\documentclass[11pt]{article}
\usepackage[letterpaper, margin=1in]{geometry}
\usepackage{amsmath, amssymb, amsthm}
\usepackage{mathtools}
\usepackage{bbm}
\usepackage{graphicx}
\usepackage{url}
\usepackage{hyperref}
\usepackage[capitalise]{cleveref}
\usepackage{algorithm}
\usepackage[noend]{algpseudocode}
\usepackage{subcaption}
\usepackage{xspace}
\usepackage{mathrsfs}
\usepackage{xcolor}
\usepackage{tabularx}
\usepackage{booktabs}
\usepackage{tikz}
\usetikzlibrary{decorations.pathreplacing}
\tikzset{
  exchange vertex/.style={
    circle,draw,fill=white,minimum size=8mm,
    inner sep=1pt,line width=.6pt
  },
  M edge/.style={
    draw=black,solid,line width=.8pt
  },
  sigma edge/.style={
    draw=black,dash pattern=on 4pt off 2.5pt,line width=.8pt
  },
  shortcut edge/.style={
    sigma edge,line width=1.1pt
  },
  row label/.style={
    font=\small\sffamily,anchor=east
  },
  note/.style={
    font=\footnotesize,align=center
  }
}

\newtheorem{theorem}{Theorem}[section]

\newtheorem{lemma}[theorem]{Lemma}

\newtheorem{claim}[theorem]{Claim}

\crefname{claim}{Claim}{Claims}
\Crefname{claim}{Claim}{Claims}
\crefname{lemma}{Lemma}{Lemmas}
\Crefname{lemma}{Lemma}{Lemmas}

\newenvironment{proof-sketch}{\noindent{\bf Sketch of Proof}\hspace*{1em}}{\qed\bigskip}
\newenvironment{proof-idea}{\noindent{\bf Proof Idea}\hspace*{1em}}{\qed\bigskip}
\newenvironment{proof-of-lemma}[1]{\noindent{\bf Proof of Lemma #1}\hspace*{1em}}{\qed\bigskip}
\newenvironment{proof-attempt}{\noindent{\bf Proof Attempt}\hspace*{1em}}{\qed\bigskip}

\makeatletter
\def\fnum@figure{{\bf Figure \thefigure}}
\def\fnum@table{{\bf Table \thetable}}
\long\def\@mycaption#1[#2]#3{\addcontentsline{\csname
  ext@#1\endcsname}{#1}{\protect\numberline{\csname
  the#1\endcsname}{\ignorespaces #2}}\par
  \begingroup
    \@parboxrestore
    \small
    \@makecaption{\csname fnum@#1\endcsname}{\ignorespaces #3}\par
  \endgroup}
\def\mycaption{\refstepcounter\@captype \@dblarg{\@mycaption\@captype}}
\makeatother

\newcommand{\mathify}[1]{\ifmmode{#1}\else\mbox{$#1$}\fi}
\newcommand{\bigO}O

\newcommand{\remove}[1]{}
\newcommand{\ignore}[1]{}

\def\poly{{\rm poly}}

\newtheorem{property}[theorem]{Property}
\crefname{property}{Property}{Properties}
\Crefname{property}{Property}{Properties}
\usepackage{enumitem}
\usepackage{needspace}
\hypersetup{
  colorlinks=true,
  linkcolor=black,
  filecolor=blue,
  urlcolor=blue,
  citecolor=blue,
  pdfpagemode=FullScreen,
}

\usepackage[sortcites, natbib=true, url=false, style=alphabetic, doi=false, isbn=false, maxbibnames=9, maxalphanames=4,minalphanames=3]{biblatex}
\AtEveryBibitem{%
  \clearfield{note}
  \clearname{editor}
  \clearlist{language}
  \clearfield{location}
}

\IfFileExists{references.bib}{\addbibresource{references.bib}}{}
\renewcommand{\vec}[1]{\mathbf{#1}}
\newcommand{\one}{\vec{1}}
\newcommand{\U}{\mathcal{U}}

\newcommand{\ALG}{\mathrm{ALG}}
\newcommand{\ADV}{\mathrm{ADV}}
\newcommand{\E}{\mathbb{E}}

\newcommand{\Harmonic}{\textnormal{\textsc{Harmonic}}\xspace}

\title{The \Harmonic Online $k$-Server Algorithm Is $\poly(k)$-Competitive}
\author{Yichen Huang\thanks{Supported in part by NSF grant CNS-2107078. Email: \texttt{yichenhuang@g.harvard.edu}.}\\Harvard University}
\date{\today}

\begin{document}
\maketitle

\begin{abstract}
The online $k$-server problem is one of the most fundamental problems in online algorithms. In this problem, $k$ servers are located in a metric space and serve a sequence of requests at points in the metric space by moving a server there. The \Harmonic algorithm is a very natural algorithm: it chooses a server with probability inversely proportional to its distance from the request. It is also highly efficient, being memoryless and answering each request in $O(k)$ time. However, its competitive ratio has remained unsettled between the $O(2^k \log k)$ upper bound (Bartal and Grove, JACM 2000) and the $k(k+1)/2$ lower bound (Raghavan and Snir, 1994, IBM J. Res. Dev.). We prove that \Harmonic is $2k^4$-competitive on every metric space against adaptive online adversaries. Our analysis uses a potential combining corrected electrical energy with the distances of online servers to their assigned adversary locations.
\end{abstract}

\pagenumbering{gobble}
\newpage
\pagenumbering{arabic}

\section{Introduction}\label{sec:introduction}
The $k$-server problem is one of the central problems in online algorithms~\cite{MMS90,KP95,BCLLM18,BCR23,CKZ26}. In this problem, $k$ servers occupy points in a metric space $(V,d)$. At each time $t$, a request appears at a point $r_t\in V$, and the algorithm must move a server to $r_t$. The cost is the total distance traveled by the servers. An online algorithm must select a server without knowledge of future requests. The goal is to minimize the competitive ratio, which compares the algorithm's cost with that of an optimal schedule that knows the requests in advance. 

The work function algorithm (WFA) was recently shown to achieve competitive ratio $k$~\cite{CKZ26}, which is optimal both for deterministic algorithms and for randomized algorithms against adaptive online adversaries~\cite{MMS90,RS94}. Against oblivious adversaries, a recent preprint establishes an $O(\log^2 k)$ upper bound~\cite{OAI26}, matching the lower bound in~\cite{BCR23}.

Although information-theoretically optimal, these competitive ratios come with heavy computational costs. Known implementations of WFA answer each request in time that either depends exponentially on $k$ or grows with the number of previous requests~\cite{CC26}, and they require comparable memory to store the work-function values or prior request locations~\cite{RS22WFA}. The implementation of the $O(\log^2 k)$-competitive algorithm also has time and memory usage that grow with the request count~\cite{OAI26Uniform}. When time and space are limited, we would like a fast algorithm with small memory usage.

A particularly attractive candidate in this regard is the \Harmonic algorithm, introduced in~\cite{RS89,RS94}. On an uncovered request, it selects a server for the request with probability inversely proportional to its distance from the request. This rule is simple and intuitive: nearby servers are more likely to move, while randomization allows every server to be selected. It is also memoryless,\footnote{A randomized server algorithm is memoryless if the distribution of its next move depends only on its current server configuration and the current request (cf. \cref{sec:preliminaries}).} and each decision takes $O(k)$ distance evaluations and arithmetic operations, independently of the number of points in the metric and the length of the request history.


Unfortunately, the competitive ratio of the \Harmonic algorithm has remained poorly understood. \cite{RS94} proves a lower bound of $k(k+1)/2$, even against an oblivious adversary. \cite{Grove91} establishes that \Harmonic is competitive, and \cite{BG00} gives the best previously known general upper bound, $O(2^k\log k)$, against adaptive online adversaries. The gap between these bounds leaves open whether \Harmonic's simplicity incurs a polynomial or an exponential loss.

We prove the first polynomial upper bound on \Harmonic's competitive ratio. Thus \Harmonic has a competitive ratio and per-request response time that are both polynomial in $k$ and independent of the metric size and the request history. To the best of our knowledge, this is the first such joint guarantee.

\begin{theorem}\label{thm:main}
\Harmonic is $2k^4$-competitive for the online $k$-server problem on every metric space against adaptive online adversaries.
\end{theorem}

We prove~\cref{thm:main} via a novel potential function based on electrical
networks. Electrical networks have been used to analyze how \Harmonic
repairs a single mismatch while the adversary stays fixed~\cite{BCNR02}.
However, this analysis does not directly extend to general adaptive
interactions: several discrepancies may coexist, and the adversary may
move again before they are repaired.

To account for these mismatches together, we assign each online server to
an adversary location. The assignment need not be injective: several
servers may share an assigned location. An assignment is charged both the total assignment
distance and the resulting imbalance relative to one assigned server per
adversary location. The potential is the minimum of a weighted sum of these two charges
over all assignments.

To measure the imbalance, we build an electrical network on the adversary
locations, joining every pair by a resistor whose resistance is their
distance. Each location with extra assigned servers supplies one unit of
current per extra server, while each location with no assigned server
receives one unit. Among flows meeting these supplies and demands, the
electrical flow minimizes the energy: the sum of resistance times squared
current.

Energy alone, however, need not decrease in expectation when \Harmonic{}
moves. Because it is symmetric, the energy loses information about \emph{direction}: swapping an extra server
and a missing one preserves their energy, although serving distinguishes
the source from the requested destination. To capture this distinction,
we add a linear correction consisting of a location-dependent charge per
assigned server.

We choose this correction so that the decrease in corrected energy from
each move between assigned locations is proportional to a voltage
difference in one common electrical field. Dividing by the assigned
movement distance then gives a quantity proportional to current, matching
\Harmonic's reciprocal-distance probabilities. Kirchhoff's law and the
triangle inequality control the resulting weighted currents, while the
assignment-distance charge accounts for the difference between assigned
and actual movements. Together, these bounds show that the expected
potential decrease pays for \Harmonic's expected movement. Finally,
bounding current magnitudes by the total supply also controls sensitivity
to adversary movement, limiting the potential increase to $2k^4$ times
the distance moved.

\paragraph{Organization.} \Cref{sec:preliminaries} sets up the potential framework, and \cref{sec:warmup} reviews the single-mismatch case. \Cref{sec:motivation} defines and motivates the potential and establishes its properties. \Cref{sec:potential} combines these properties to prove the competitive ratio. Omitted proof details and examples can be found in the appendices.



\subsection{Related Work}
For \Harmonic itself, \cite{CL92} proves the tight ratio $3$ for two servers; \cite{CS00} gives a simpler proof. \cite{BCNR02} proves the conjectured $k(k+1)/2$ bound against a restricted adversary. After moving one server from a configuration shared with \Harmonic, this adversary stays fixed and repeatedly requests an adversary-occupied point not covered by \Harmonic until the two configurations agree again.

Electrical serving rules also appear in related memoryless algorithms. When distances can be realized as effective resistances, the \textsc{RWalk} algorithm in~\cite{CDRS93} computes the corresponding conductances on the current server locations and request, and uses them to select the serving server. In the hard $k$-taxi problem, requests specify pickup and drop-off points, and only empty travel is charged. On hierarchically separated trees, the \textsc{Flow} algorithm in~\cite{CK19} chooses a taxi according to the fraction of electrical current reaching it.

The weighted $k$-server problem assigns each server its own movement cost per unit distance. On uniform metrics, \cite{CV13} develops a linear-system potential analysis of randomized memoryless algorithms, including an analogue of \Harmonic that samples servers inversely proportionally to their weights.

\section{Preliminaries}\label{sec:preliminaries}

Let $(V,d)$ be a metric space and $[k]=\{1,\ldots,k\}$. We assume $k\ge2$, since the case $k=1$ is trivial. A \emph{configuration} is a multiset of $k$ points of $V$, allowing several servers at the same point. For bookkeeping, we label the servers and represent the online and adversary configurations by tuples $X=(x_1,\ldots,x_k)$ and $Y=(y_1,\ldots,y_k)$. A request at $r\in V$ is \emph{covered} by a configuration if it contains a server at $r$. The total cost is the sum of all server movement distances.

On a covered request, \Harmonic leaves its configuration unchanged. On an uncovered request at $r$, it samples a label $i\in[k]$ with probability
\[
 p_i=\frac{1}{Z_r\,d(x_i,r)},
 \qquad Z_r=\sum_{j=1}^k\frac{1}{d(x_j,r)},
\]
and moves server $i$ to $r$.

We compare \Harmonic with an \emph{adaptive online adversary}, which generates the requests and also serves them. The adversary may choose each request and its serving moves based on the algorithm's past actions, but must serve the request before the algorithm makes its fresh random choice. An algorithm is $\rho$-competitive if, for every adaptive online adversary and every finite number of rounds, its expected total cost is at most $\rho$ times the expected adversary cost, plus an additive constant depending only on the metric and initial configurations.

\paragraph{Distinct Adversary Locations.}
We may assume that the adversary is \emph{lazy}: it moves exactly one server directly to a request when that point is uncovered by its own configuration, and otherwise does not move. The standard lazification reduction is causal and does not increase cost~\cite[Lemma~1]{MMS90}, so it also applies to an adaptive online adversary. We may also take its initial locations to be distinct.\footnote{If initial adversary servers coincide, attach a private edge of length $\varepsilon>0$ at each server's original location and start that server at the new endpoint. Keep all requests and initial online positions in the original metric, so \Harmonic's probabilities and costs are unchanged. Any original adversary schedule lifts with at most $k\varepsilon$ extra cost and can then be lazified. This fixed extra cost affects only the additive constant.} Every lazy move then goes to a previously unoccupied point, preserving distinctness. Henceforth, the adversary is lazy and its locations $y_1,\ldots,y_k$ are always pairwise distinct.

\paragraph{Potential Function Framework.}
For a randomized server algorithm $\ALG$, we use a finite, nonnegative potential $\Phi(X,Y)$ as credit for future online movement. Written in terms of labeled tuples, $\Phi:V^k\times V^k\to\mathbb R_{\ge0}$ is invariant under permutations of either tuple, so its value depends only on the two configurations.

Analyze each round in two stages: the adversary first covers the request, then $\ALG$ serves it with the adversary configuration held fixed. For a parameter $\rho\ge1$, we seek the following two properties for the potential.

\begin{property}[Adversary Movement]\label[property]{prop:adversary-movement}
For every online configuration $X$, if one adversary server moves a distance $\delta$ from configuration $Y$ to $Y'$, then
\[
  \Phi(X,Y')-\Phi(X,Y)\le\rho\delta.
\]
\end{property}

\begin{property}[Online Movement]\label[property]{prop:online-movement}
Condition on any complete past, request $r$, and adversary serving moves. Let $X,Y$ be the current configurations, with $Y$ covering $r$. If $X'$ is the random configuration after $\ALG$ serves $r$ and $c$ is its total movement cost, then
\[
  \E\bigl[\Phi(X,Y)-\Phi(X',Y)\bigr]\ge\E[c].
\]
\end{property}

These two properties suffice to prove competitiveness, as the following lemma shows.

\begin{lemma}[Potential Criterion]\label[lemma]{lem:potential-criterion}
Under the adversary conventions above, a randomized server algorithm admitting a finite, nonnegative potential satisfying \cref{prop:adversary-movement,prop:online-movement} for some $\rho\ge1$ is $\rho$-competitive, with additive constant $\Phi(X_0,Y_0)$ for initial configurations $X_0,Y_0$.
\end{lemma}
\begin{proof}
Let $\ALG_t,\ADV_t$ be the online and adversary costs in round $t$, and let $X_t,Y_t$ be their configurations at its end. The adversary first changes $Y_{t-1}$ to $Y_t$ while the online configuration remains $X_{t-1}$. Applying \cref{prop:adversary-movement} to this stage and \cref{prop:online-movement} to the subsequent online stage gives
\[
  \E\bigl[\ALG_t+\Phi(X_t,Y_t)-\Phi(X_{t-1},Y_{t-1})\bigr]\le\rho\E[\ADV_t].
\]
Summing over $N$ rounds telescopes the potential changes. Nonnegativity of the final potential yields
\[
  \E\left[\sum_{t=1}^N \ALG_t\right]\le\rho\E\left[\sum_{t=1}^N \ADV_t\right]+\Phi(X_0,Y_0).
\]
The finite initial potential is independent of the requests, giving the required competitive bound.
\end{proof}

We apply this criterion to \Harmonic with $\rho=2k^4$. Covered requests require no movement and satisfy \cref{prop:online-movement} immediately. For an uncovered request at $r$, the expected movement cost is $\sum_i p_i d(x_i,r)=k/Z_r$. Write $X^{i\to r}$ for the configuration after server $i$ moves to $r$. Substituting the sampling probabilities shows that \cref{prop:online-movement} is equivalent to
\[
  \sum_{i=1}^k\frac{\Phi(X,Y)-\Phi(X^{i\to r},Y)}{d(x_i,r)}\ge k,
\]
whenever $Y$ covers $r$ and $X$ does not. Dividing by $Z_r$ gives an expected decrease of at least $k/Z_r$, as required. We construct the potential in Section~\ref{sec:motivation} and verify the two properties in Section~\ref{sec:potential}.

\section{Warmup: A Single Hole and Its Electrical Interpretation}\label{sec:warmup}
We recall the classical moving-hole argument~\cite[Section~3]{BCNR02}. It gives an exact potential for repairing one mismatch with a fixed adversary and explains the electrical interpretation of a random walk. Our potential will be a substantial generalization of this electrical interpretation. Readers familiar with this correspondence may skip to \cref{sec:motivation}.

\paragraph{A Walk That Repairs One Mismatch.}
Suppose each configuration occupies $k$ distinct points and they agree at $k-1$ of them. Label their union $z_1,\ldots,z_{k+1}$, with $z_a$ occupied only online and $z_s$ only by the adversary. The unique point without an online server is the \emph{hole}, initially $z_s$. Hold the adversary fixed and repeatedly request the hole. Each serving move fills it and creates a new hole at the server's old position. The process stops when the hole reaches $z_a$, since the configurations then agree. Until then, every request is already covered by the adversary.

Write $d_{ij}=d(z_i,z_j)$. At a hole $z_i\ne z_a$, every other point has one online server, so the hole jumps to $z_j$ with probability $1/(Z_i d_{ij})$, where $Z_i=\sum_{j\ne i}1/d_{ij}$. Each jump has the same length as its serving move. Let $\ell_a(i)$ be the expected remaining walking distance until reaching $z_a$, with $\ell_a(a)=0$. These expectations are finite because the finite walk has a uniformly positive chance of reaching $z_a$ at every step and bounded step lengths. Conditioning on the next jump and clearing the normalizing factor gives
\[
  \sum_{j\ne i}\frac{\ell_a(i)-\ell_a(j)}{d_{ij}}=k\qquad(i\ne a).
\]
The right side counts $k$ possible jumps, each contributing $d_{ij}/d_{ij}=1$ to the movement cost. Dividing by $Z_i$ shows that the expected decrease of $\ell_a$ equals the expected serving cost $k/Z_i$.

\paragraph{Reading the Identity as an Electrical Balance.}
Join every pair of points by an edge with resistance $d_{ij}$. A \emph{voltage} is a real value at a vertex; the signed \emph{current} on an edge is the voltage difference divided by its resistance. Set
\[
  h_i=\frac{\ell_a(i)}k,\qquad F_{ij}=\frac{h_i-h_j}{d_{ij}}\quad(i\ne j),\qquad F_{ii}=0.
\]
The first-step identity says that every vertex other than $a$ supplies one unit of current. Since $F_{ij}=-F_{ji}$, these supplies must all reach $a$:
\[
  \sum_jF_{ij}=1\quad(i\ne a),\qquad \sum_jF_{aj}=-k.
\]
The currents are signed quantities derived from the expected cost, while the walk chooses destinations by reciprocal distance. The connection is that dividing a potential difference by a serving distance gives $k$ times the corresponding current. This is the classical random-walk interpretation~\cite[Section~1.3]{DS84}.

\medskip

The warmup's cost-to-go relies on one hole and one fixed absorbing point. Several holes cannot be described by the same one-hole walk, and adversary movement can change the network and the repair target.

To retain the electrical structure, reverse the warmup field and divide its currents by $k+1$. Label $a$ then supplies $k/(k+1)$ and every other label receives $1/(k+1)$ in net. Starting with a unit at $a$, this leaves $1/(k+1)$ there and gives the same share to every other label. In the general case we first assign online servers to the current adversary locations, apply the analogous construction on their common $k$-label network, and superpose one redistribution flow per assigned server. 

\section{The Potential and Its Motivation}\label{sec:motivation}
Fix an adversary configuration $Y=(y_1,\ldots,y_k)$ and write $d_{ij}=d(y_i,y_j)$. Our potential assigns each online server to an adversary label, evaluates a corrected electrical potential for the assigned counts, and adds credit for the assignment distances. Its value is the minimum of this combined quantity over all assignments. We give the complete definition first, then explain why its components are needed and establish their properties.

An assignment is a function $f:[k]\to[k]$ sending server $\ell$ at $x_\ell$ to label $f(\ell)$. Its occupancy vector $\vec{q}_f$ has coordinates $q_{f,i}=|f^{-1}(i)|$. More generally, an occupancy vector $\vec{q}\in\mathbb Z_{\ge0}^k$ gives counts of total mass $\sum_iq_i=k$. Write $\one=(1,\ldots,1)$ and let $\vec{e}_i$ be the $i$th coordinate vector.

The zero-sum vector $\vec{q}-\one$ describes the imbalance of these counts: positive entries specify net supply and negative entries net demand. For any zero-sum vector $\vec{a}$, define its \emph{electrical energy} by minimizing over signed flows $F_{ij}=-F_{ji}$ with $F_{ii}=0$:
\[
  E_Y(\vec{a})\coloneqq\min_{\sum_jF_{ij}=a_i}\sum_{i<j}d_{ij}F_{ij}^2.
\]

Let $c_i\coloneqq E_Y(\vec{e}_i-\one/k)$ be the energy of redistributing one unit at label $i$ uniformly among all labels, and define the corrected occupancy potential
\[
  \U_Y(\vec{q})\coloneqq\frac{k^2}{2}\left[E_Y(\vec{q}-\one)+\sum_iq_ic_i\right].
\]
The full potential is
\[
  \Phi(X,Y)\coloneqq\min_{f:[k]\to[k]}\left\{\U_Y(\vec{q}_f)+k^3\sum_{\ell=1}^k d(x_\ell,y_{f(\ell)})\right\}.
\]

The energy measures the joint imbalance from one server per adversary label. The linear correction expresses the decrease in the occupancy term caused by each movement as a voltage difference, linking it to \Harmonic's sampling probabilities. The distance term then transfers this calculation to actual server positions. We next motivate these three components gradually.

Section~\ref{sec:energy} explains why energy is useful and why it alone can fail. Section~\ref{sec:correction} derives the corrected serving identity. Section~\ref{sec:sensitivity} bounds sensitivity and justifies the assignment-distance credit. Section~\ref{sec:potential} combines these ingredients to prove the competitive guarantee.

\subsection{The Electrical Energy}\label{sec:energy}
For this subsection and the next, temporarily place each online server at its assigned adversary location and hold $Y$ fixed. We study the occupancy term $\U_Y(\vec{q})$ on these assigned counts. Several actual servers at distinct points may share an assigned label, so some labels can have extra units while others have none.

The difference $\vec{q}-\one$ describes the excesses and shortages. For example, $\vec{q}=(2,2,0,0)$ gives $(1,1,-1,-1)$: one extra server at each of the first two labels and one missing at each of the last two.

For a general zero-sum vector $\vec{a}$, a label with $a_i>0$ supplies $a_i$ units, while a label with $a_i<0$ receives $-a_i$ units. The signed current $F_{ij}$ is the amount sent from $i$ to $j$; a negative value means flow in the reverse direction. A flow is feasible for $\vec{a}$ when $\sum_jF_{ij}=a_i$ at every label, routing the supplies to the receiving labels. The network has one edge of resistance $d_{ij}$ between each pair of labels, and minimizing the quadratic cost gives its electrical flow. When $Y$ is fixed, we omit the subscript on $E_Y$.

The following standard fact explains the benefit of the quadratic cost. A minimizing flow can be described by one real value $h_i$ at each label. These values are called \emph{voltages}; their differences determine the flow on every positive-length edge.

\begin{lemma}[Electrical Characterization and Expansion]\label[lemma]{lem:flow}
For every zero-sum vector $\vec{a}$, the minimum defining $E(\vec{a})$ is attained. A feasible flow $F$ is minimizing if and only if there are real values $h_1,\ldots,h_k$ satisfying
\[
  d_{ij}F_{ij}=h_i-h_j\qquad(i\ne j).
\]
For such a flow and every zero-sum vector $\vec{b}$,
\[
  E(\vec{a}+\vec{b})=E(\vec{a})+2\sum_i b_i h_i+E(\vec{b}).
\]
\end{lemma}

This is a standard minimum-energy characterization~\cite[Section~1.3.5]{DS84}; Appendix~\ref{app:electrical} gives a proof. Since the adversary locations are distinct, $F_{ij}=(h_i-h_j)/d_{ij}$ for every $i\ne j$.

The superposition promised in the warmup takes place on these $k$ labels. For a server assigned to label $i$, the analogous reversed, normalized field has excess $\vec{e}_i-\one/k$: label $i$ supplies $1-1/k$, while every other label receives $1/k$ in net. Adding one such flow per assigned server gives total excess
\[
  \sum_iq_i(\vec{e}_i-\one/k)=\vec{q}-\one.
\]
Voltages and currents add on the same network, so \cref{lem:flow} identifies the sum as the minimizing flow for this joint imbalance. Its quadratic energy includes the interactions among all these flows.

Energy is nonnegative, $E(\vec{0})=0$, and reversing all flows shows that $E(-\vec{a})=E(\vec{a})$. Thus $E(\vec{q}-\one)$ measures how much imbalance remains. The example on a uniform metric explains its attraction as (part of) a potential.

\paragraph{Energy Pays on a Uniform Metric.}
Suppose the $k$ adversary locations are distinct and all distances between them are $1$. Taking $h_i=(q_i-1)/k$ in \cref{lem:flow} gives
\[
  E(\vec{q}-\one)=\frac1k\sum_i(q_i-1)^2.
\]
Request an empty label $s$. If a server moves from an occupied label $i$ to $s$, only these two coordinates change, so the energy decrease is
\[
  E(\vec{q}-\one)-E(\vec{q}-\vec{e}_i+\vec{e}_s-\one) =\frac{(q_i-1)^2+1-(q_i-2)^2}{k} =\frac{2(q_i-1)}k.
\]
Serving from a label with one server leaves energy unchanged. Serving from a label with at least two servers removes one extra server and one hole, decreasing the imbalance.

Every server is at distance $1$ from the request, so \Harmonic selects each with probability $1/k$, and selects label $i$ with probability $q_i/k$. An empty label forces some other label to have at least two servers, giving
\[
  \E[\text{energy decrease}] =\frac{2}{k^2}\sum_iq_i(q_i-1)\ge\frac4{k^2}.
\]
Each serving move costs $1$. Thus $(k^2/4)E(\vec{q}-\one)$ pays the online cost in expectation with the adversary fixed, even when several holes coexist. 

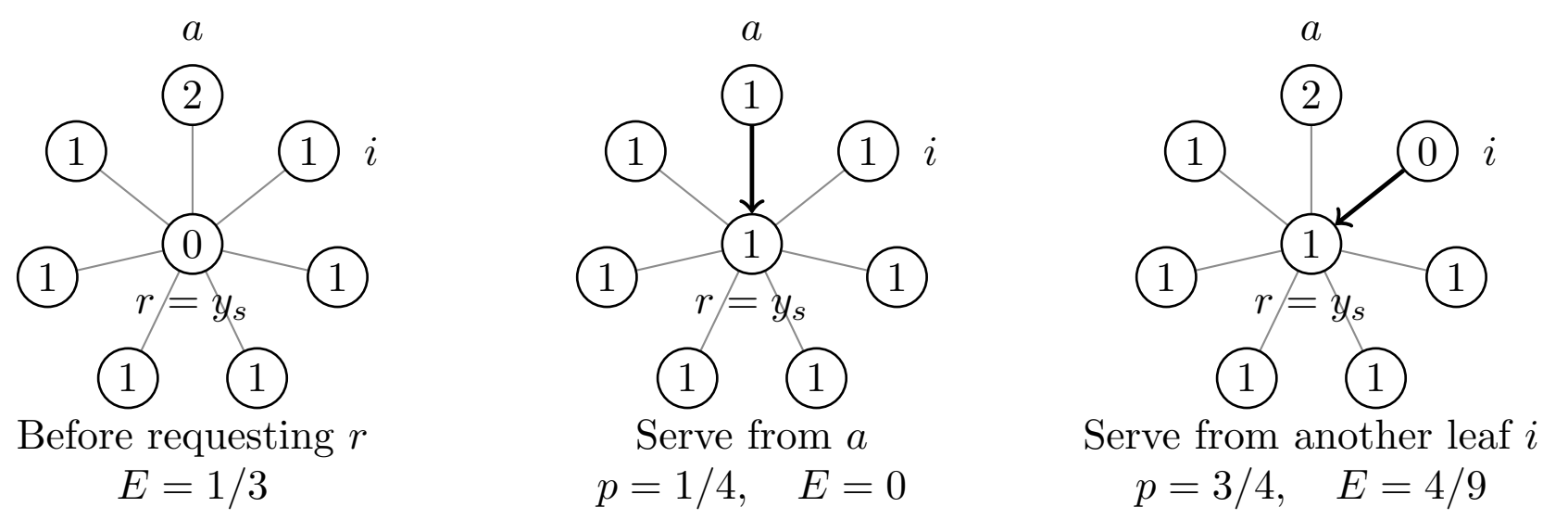
\begin{figure}[tbp]
\centering
\begin{tikzpicture}[scale=0.85,transform shape,font=\small,
  star vertex/.style={circle,draw,fill=white,minimum size=5mm,inner sep=0pt,line width=.5pt}]
  \foreach \panel/\centerq/\aq/\iq in {0/0/2/1,1/1/1/1,2/1/2/0} {
    \begin{scope}[xshift={4.7cm*\panel}]
      \foreach \j in {0,...,6} {
        \pgfmathsetmacro{\angle}{90-360*\j/7}
        \draw[black!45] (0,0)--(\angle:1.25);
      }
      \node[star vertex] (center) at (0,0) {$\centerq$};
      \foreach \j in {0,...,6} {
        \pgfmathsetmacro{\angle}{90-360*\j/7}
        \ifnum\j=0\def\occupancy{\aq}\else
          \ifnum\j=1\def\occupancy{\iq}\else\def\occupancy{1}\fi
        \fi
        \node[star vertex] (leaf\j) at (\angle:1.25) {$\occupancy$};
      }
      \node[above=2pt] at (leaf0.north) {$a$};
      \node[right=2pt] at (leaf1.east) {$i$};
      \node[below=1pt] at (center.south) {$r=y_s$};
      \ifnum\panel=0
        \node[align=center] at (0,-1.85) {Before requesting $r$\\$E=1/3$};
      \fi
      \ifnum\panel=1
        \draw[->,line width=.9pt] (leaf0)--(center);
        \node[align=center] at (0,-1.85) {Serve from $a$\\$p=1/4,\quad E=0$};
      \fi
      \ifnum\panel=2
        \draw[->,line width=.9pt] (leaf1)--(center);
        \node[align=center] at (0,-1.85) {Serve from another leaf $i$\\$p=3/4,\quad E=4/9$};
      \fi
    \end{scope}
  }
\end{tikzpicture}
\caption{One adversary server occupies each vertex; numbers give online occupancies. The star specifies the metric, while the electrical network has an edge between every pair of vertices. The last panel represents any of the six leaves other than $a$.}
\label{fig:star-energy}
\end{figure}

\paragraph{Limitations of the Energy.} Energy alone can fail on a nonuniform metric. For example, with $k=8$, take the star metric in \cref{fig:star-energy}, with seven unit-length leaves, an empty center, and leaf counts $(2,1,\ldots,1)$. Requesting the center changes the energy from $1/3$ to $0$ with probability $1/4$ and to $4/9$ with probability $3/4$. Its expected decrease is therefore $1/3-(3/4)(4/9)=0$, although every move costs $1$; Appendix~\ref{app:star} gives the full calculation.

The main problem is that the energy, due to its symmetry, disregards the \emph{direction} of the movement. In particular, energy treats an extra server at $a$ and a hole at $s$ the same as the reversed roles:
\[
  E(\vec{e}_a-\vec{e}_s)=E(\vec{e}_s-\vec{e}_a).
\]
Serving has a direction, however: the server moves into the requested hole. Swapping the extra server and the hole can change the serving probabilities and costs, even though the energy is unchanged. This issue was absent on the uniform metric because every label has the same geometry. The linear term $\sum_iq_ic_i$ in $\U_Y$ supplies this missing dependence on the roles of the locations. We next explain why these particular coefficients give the serving identity we need.

\subsection{Deriving the Corrected Potential}\label{sec:correction}
The expected potential decrease averages the decreases from individual server movements, with weights inversely proportional to their distances. We therefore want the decrease from each movement to be a voltage difference: dividing by distance gives current, so the flow equations and metric inequalities can control all choices together. The linear correction produces this form for one common flow, as we now show.

Return to a general occupancy $\vec{q}$ and request label $s$. A serving move from label $i$ changes the counts to $\vec{q}-\vec{e}_i+\vec{e}_s$. We temporarily add a server at $s$ before choosing which old server to remove. 

The temporary configuration $\vec{q}+\vec{e}_s$ has $k+1$ servers. Dividing this mass equally among the $k$ labels gives the target count $1+1/k$, so its excess relative to this target is
\[
  \vec{a}\coloneqq\vec{q}+\vec{e}_s-(1+1/k)\one.
\]
Let $F$ be a minimizing flow for $\vec{a}$, with voltages $\vec{h}$. Removing one server also lowers the uniform target by $1/k$ at every label. Consequently, the old and new excess vectors are
\[
  \vec{q}-\one=\vec{a}-(\vec{e}_s-\one/k),\qquad \vec{q}-\vec{e}_i+\vec{e}_s-\one=\vec{a}-(\vec{e}_i-\one/k).
\]

Both excess vectors are obtained from the same $\vec{a}$ by removing a unit-spreading vector. The quadratic expansion therefore expresses both energies using the voltages of this common flow. The linear correction cancels the remaining spreading-energy terms, giving the following identity.

\begin{claim}[Serving Move Identity]\label[claim]{clm:move-identity}
Fix an occupancy $\vec{q}$ of mass $k$ and a request label $s$. Let $F$ be a minimizing flow with excess vector $\vec{a}=\vec{q}+\vec{e}_s-(1+1/k)\one$ and voltages $\vec{h}$. For each $i$ with $q_i>0$,
\[
  \Delta_i\coloneqq\U_Y(\vec{q})-\U_Y(\vec{q}-\vec{e}_i+\vec{e}_s) =k^2(h_i-h_s).
\]
\end{claim}
\begin{proof}
The vector $\vec{e}_t-\one/k$ defining $c_t$ represents spreading one unit from label $t$ equally among all labels. Apply \cref{lem:flow} with $\vec{b}=-(\vec{e}_t-\one/k)$. Reversing a flow preserves energy, so
\[
  E_Y\bigl(\vec{a}-(\vec{e}_t-\one/k)\bigr)=E_Y(\vec{a})-2h_t+\frac2k\sum_jh_j+c_t.
\]
Taking $t=s$ gives the old occupancy's energy, while $t=i$ gives the serving outcome's energy. Subtracting cancels the common terms:
\[
  E_Y(\vec{q}-\one)-E_Y(\vec{q}-\vec{e}_i+\vec{e}_s-\one)=2(h_i-h_s)+c_s-c_i.
\]
The linear term decreases by $c_i-c_s$, canceling the extra $c_s-c_i$. Multiplying the remaining $2(h_i-h_s)$ by $k^2/2$ proves the identity.
\end{proof}
\paragraph{Remark (Charge Interpretation).}
Let $G$ be the pseudoinverse of the network Laplacian with conductances $1/d_{ij}$, so $G\one=\vec{0}$. By \cref{lem:flow}, $E_Y(\vec{q}-\one)=\vec{q}^{\top}G\vec{q}$ and $c_i=G_{ii}$, giving
\[
  \U_Y(\vec{q}_f)=k^2\sum_{1\le\ell\le m\le k}G_{f(\ell),f(m)}.
\]
This is the interaction energy of assigned unit charges, counting each unordered pair and each self-interaction once; in this form, \cref{clm:move-identity} expresses ``work equals charge times potential difference'' in the common field $G(\vec{q}+\vec{e}_s)$, with overall scale $k^2$.

For $d_{is}>0$, division by assigned serving distance gives $\Delta_i/d_{is}=k^2F_{is}$. Thus the reciprocal-distance weights turn the corrected potential decrease into current. Section~\ref{sec:potential} will use this relation to bound the expected decrease; first we must account for the difference between assigned and actual serving distances.



\subsection{Sensitivity and Assignment Credit}\label{sec:sensitivity}
We now return to the actual online positions. The preceding calculation treats servers at their assigned locations; the distance credit must bridge the gap between those positions and $X$. A server at $x$ assigned to $y_i$ serves a request $r=y_s$ at distance $d(x,r)$, whereas its assigned move has distance $d_{is}$. The triangle inequality bounds the difference by $d(x,y_i)$. To compensate for this discrepancy, we need an upper bound on the magnitude of an occupancy change per unit distance; individual decreases can have either sign.

The serving identity reduces this task to bounding voltage differences. We first prove this voltage bound together with an energy sensitivity bound. The latter will also be used in Section~\ref{sec:potential} to charge the potential increase caused by adversary movement.

\begin{lemma}[Voltage and Energy Stability]\label[lemma]{lem:stability}
Fix $Y$ and a zero-sum vector $\vec{a}$, and put $T=\sum_i\max\{a_i,0\}$. Then, the voltages $\vec h$ of any minimizing flow of $\vec{a}$ satisfy \[|h_i-h_j|\le T d_{ij}.\] If $Y'$ is obtained from $Y$ by moving one adversary label a distance $\delta$, then \[E_{Y'}(\vec{a})\le E_Y(\vec{a})+2T^2\delta.\]
\end{lemma}
\begin{proof}
Choose a minimizing flow $F$ with voltages $\vec{h}$, and orient every nonzero edge in the direction of its positive flow. Since the adversary locations are distinct, voltage strictly decreases along every such edge. Hence there are no directed cycles.

Order the labels so that every positive edge goes from an earlier label to a later one. Such an order exists by repeatedly removing a label with no incoming edge. At any cut in this order, all crossing flow goes forward. Its total is the sum of the excesses before the cut, which is at most $T$. The cut just before a label bounds its total incoming flow by $T$; the cut just after it bounds its total outgoing flow by $T$. It follows that
\[
  |F_{ij}|\le T,\qquad \sum_{j\ne i}F_{ij}^2\le T\sum_{j\ne i}|F_{ij}|\le2T^2.
\]
The first inequality and $h_i-h_j=d_{ij}F_{ij}$ give the voltage bound. Moving label $i$ changes only its incident edge lengths, each by at most $\delta$. The same flow is feasible after the move, and its energy increases by at most $\delta\sum_{j\ne i}F_{ij}^2\le2T^2\delta$. The new minimum is no larger than this candidate energy.
\end{proof}

We can now apply the voltage bound to the common flow from the serving identity. Its total positive excess is at most $k$, so multiplying voltage differences by the identity's factor $k^2$ gives the following $k^3$ bound.

\begin{lemma}[Individual Move Bound]\label[lemma]{lem:move-bound}
For any adversary configuration $Y$, occupancy $\vec{q}\in\mathbb Z_{\ge0}^k$ of mass $k$, and labels $i,s$ with $q_i>0$,
\[
  \bigl|\U_Y(\vec{q})-\U_Y(\vec{q}-\vec{e}_i+\vec{e}_s)\bigr|\le k^3d_{is}.
\]
\end{lemma}
\begin{proof}
Use the common flow from \cref{clm:move-identity}, whose excess vector is $\vec{a}=\vec{q}+\vec{e}_s-(1+1/k)\one$. Every coordinate satisfies $a_j\le q_j$, so the total positive excess is at most $k$. The voltage bound in \cref{lem:stability} and the serving identity in \cref{clm:move-identity} give
\[
  \bigl|\U_Y(\vec{q})-\U_Y(\vec{q}-\vec{e}_i+\vec{e}_s)\bigr|=k^2|h_i-h_s|\le k^3d_{is}.\qedhere
\]
\end{proof}

This bound explains the coefficient in $\Phi$. When a server serves $r$ and is assigned to label $s$ afterward, its old assignment distance disappears. The resulting credit $k^3d(x,y_i)$ is large enough to absorb the discrepancy between actual and assigned serving distances. Section~\ref{sec:potential} makes this comparison precise and proves that the total expected decrease pays for the online cost.

\section{Competitive Analysis}\label{sec:potential}


We verify \cref{prop:adversary-movement,prop:online-movement} for $\Phi$, taking $\rho=2k^4$. We follow the order of a round: adversary movement first, then online movement with the resulting adversary configuration held fixed.

For each comparison, we start with an assignment attaining the old minimum and choose a candidate assignment afterward. The new minimum is no larger than this candidate value. Consequently, the candidate's increase bounds the actual increase from above, and its decrease bounds the actual decrease from below.

\begin{proof}[Proof of \cref{prop:adversary-movement} for $\rho=2k^4$]
Suppose one adversary label moves a distance $\delta$, from configuration $Y$ to $Y'$. Keep an assignment attaining $\Phi(X,Y)$ and use the same labels after the move. We bound separately the change in electrical energy and in assignment distances.

For the fixed assigned counts $\vec{q}$, the vector $\vec{q}-\one$ has total positive excess at most $k-1$: at most $k$ servers contribute, and at least one occupied label keeps its target unit. Each vector $\vec{e}_i-\one/k$ has total positive excess at most $1$. Applying the energy bound in \cref{lem:stability} to the definition of $\U_Y$, with $\sum_iq_i=k$, gives
\[
  \U_{Y'}(\vec{q})-\U_Y(\vec{q}) \le k^2\bigl((k-1)^2+k\bigr)\delta\le k^4\delta.
\]
At most $k$ servers are assigned to the moved label. Each assignment distance increases by at most $\delta$, so the distance term increases by at most another $k^4\delta$. The new minimum is no larger than this candidate value, giving
\[
  \Phi(X,Y')-\Phi(X,Y)\le2k^4\delta. \qedhere
\]
\end{proof}

\paragraph{Online Movement.}
The adversary has now covered $r$; hold its resulting configuration $Y$ fixed and choose $s$ with $y_s=r$. For assigned counts $\vec{q}$ and the common flow $F$ from \cref{clm:move-identity}, the serving analysis leads to $\sum_{i\ne s}q_iF_{is}+kq_s$: currents are counted once per assigned server, and the $kq_s$ term comes from distance credit for servers assigned to $s$. Note that even when the request is uncovered, \(q_s\) may be positive since it counts servers assigned to the requested location, whose actual positions may lie elsewhere. A lower bound of $1/k$ will suffice for \cref{prop:online-movement}.


\begin{lemma}[Weighted Current Bound]\label[lemma]{lem:weighted-current}
Fix $Y$, an occupancy $\vec{q}$ of mass $k$, and a label $s$. Let $F$ be a minimizing flow with excess vector $\vec{q}+\vec{e}_s-(1+1/k)\one$. Then
\[
  \sum_{i\ne s}q_iF_{is}+kq_s\ge\frac1k.
\]
\end{lemma}
\begin{proof}
Let $I=\sum_{i\ne s}F_{is}$. Conservation at $s$ gives $I=1/k-q_s$. We first bound the weighted sum in terms of this total incoming current.

For $i\ne s$, the flow equation gives $q_i=1+1/k+\sum_jF_{ij}$. Substituting for $q_i$ and pairing the two directions of each edge between labels other than $s$ gives
\[
  \sum_{i\ne s}q_iF_{is}=\left(1+\frac1k\right)I+\sum_{i\ne s}F_{is}^2+\sum_{\substack{i<j\\i,j\ne s}}F_{ij}(F_{is}-F_{js}).
\]
The squares are nonnegative. To control the remaining interactions, we bound each pair separately.

\begin{claim}\label[claim]{clm:current-pairing}
Every pair of distinct labels $i,j\ne s$ satisfies
\[
  F_{ij}(F_{is}-F_{js})\ge-F_{is}F_{js}.
\]
\end{claim}
\begin{proof}
    
Ohm's law gives $d_{ij}F_{ij}=d_{is}F_{is}-d_{js}F_{js}$. The triangle inequalities imply \[0\le d_{is}+d_{js}-d_{ij}\le 2 \min\{d_{is}, d_{js}\}\le2\sqrt{d_{is}d_{js}},\] and hence
\[
  d_{is}F_{is}^2+d_{js}F_{js}^2-(d_{is}+d_{js}-d_{ij})F_{is}F_{js}\ge\left(\sqrt{d_{is}}|F_{is}|-\sqrt{d_{js}}|F_{js}|\right)^2\ge0.
\]
Expanding the left side and dividing by $d_{ij}$ proves the pairwise bound.
\end{proof}

By \cref{clm:current-pairing} and the identity $2\sum_{i<j,\,i,j\ne s}F_{is}F_{js}=I^2-\sum_{i\ne s}F_{is}^2$, we obtain
\[
  \sum_{i\ne s}q_iF_{is}\ge\left(1+\frac1k\right)I-\frac{I^2}{2}+\frac32\sum_{i\ne s}F_{is}^2\ge\left(1+\frac1k\right)I-\frac{I^2}{2}.
\]

Adding $kq_s$ and substituting $I=1/k-q_s$ gives
\[
  \sum_{i\ne s}q_iF_{is}+kq_s\ge\frac1k+\frac1{2k^2}+q_s\left(k-1-\frac{q_s}{2}\right)\ge\frac1k.
\]
The last inequality follows from $0\le q_s\le k$ and $k\ge2$.
\end{proof}

We can now make the distance-credit comparison precise and sum the contributions of all serving choices.

\Needspace{6\baselineskip}
\begin{proof}[Proof of \cref{prop:online-movement}]
If $X$ covers $r$, \Harmonic does not move and the property holds. Otherwise the normalized inequality from the preliminaries is sufficient. Choose $s$ with $y_s=r$ and an assignment attaining $\Phi(X,Y)$. Let $\vec{q}$ be its counts, and let $F$ and $\vec{h}$ be the common flow and voltages from \cref{clm:move-identity}. For every occupied label $i$, write $\Delta_i=\U_Y(\vec{q})-\U_Y(\vec{q}-\vec{e}_i+\vec{e}_s)$.

If a server at $x$, assigned to label $i$, serves $r$, assign it to $s$ afterward and retain the other assignments. This candidate value decreases by $\Delta_i+k^3d(x,y_i)$, so the actual potential decrease is at least this amount.
\begin{itemize}
    \item For $i\ne s$, the assigned distance $d_{is}$ is positive. By \cref{lem:move-bound} and the triangle inequality $|d_{is}-d(x,r)|\le d(x,y_i)$,
\[
  \frac{\Delta_i+k^3d(x,y_i)}{d(x,r)}\ge\frac{\Delta_i}{d_{is}}=k^2F_{is}.
\]
Indeed, multiplying by the positive denominators reduces the inequality to $\Delta_i(d_{is}-d(x,r))+k^3d(x,y_i)d_{is}\ge0$. The two-sided bound $|\Delta_i|\le k^3d_{is}$ proves this even when $\Delta_i<0$.
\item For $i=s$, \cref{lem:move-bound} gives $\Delta_i=0$. The assignment distance is $d(x,r)$, so the decrease divided by actual movement distance is at least $k^3$.

\end{itemize}
There are $q_i$ servers assigned to each label. Summing their contributions and applying \cref{lem:weighted-current} gives
\[
  \sum_{\ell=1}^k\frac{\Phi(X,Y)-\Phi(X^{\ell\to r},Y)}{d(x_\ell,r)}\ge k^2\left(\sum_{i\ne s}q_iF_{is}+kq_s\right)\ge k.
\]
Dividing by $Z_r$ gives an expected potential decrease of at least $k/Z_r$, equal to the expected serving cost. Thus \cref{prop:online-movement} holds.
\end{proof}

\begin{proof}[Proof of \cref{thm:main}]
The potential is finite and nonnegative because it minimizes finitely many finite, nonnegative assignment values. Relabeling either configuration gives a bijection of assignments with identical values, so $\Phi$ is invariant under relabeling. The two proofs above establish \cref{prop:adversary-movement,prop:online-movement} with $\rho=2k^4$. Applying \cref{lem:potential-criterion} and the adversary reductions from \cref{sec:preliminaries} proves the theorem.
\end{proof}

\section{Conclusion and Discussion}
In this work, we prove the first polynomial in $k$ upper bound on the competitive ratio of \Harmonic. Our bound leaves a gap between $2k^4$ and the conjectured ratio $k(k+1)/2$. The $k^2$ scaling of the corrected energy, together with the voltage bound, leads to $k^3$ credit per unit of assignment distance. Under adversary movement, the corrected energy and assignment-distance terms each contribute at most $k^4$ per unit distance, yielding the bound $2k^4$. It is natural to ask whether these estimates can be sharpened and whether the approach extends to other variants, such as weighted $k$-server.

\paragraph{AI Methodology.} The results were achieved with the assistance of GPT-6 Astra. The author suggested considering a potential function involving non-injective matching between the online and offline configurations. The AI subsequently combined this idea with the electric field viewpoint and developed an initial proof of $O(k^4)$-competitiveness on ultrametrics. The author then worked iteratively with the AI to simplify the argument and generalize it to arbitrary metrics. GPT-6.1 Sol was also used to revise the exposition, organize proofs, and check intermediate calculations. The author takes full responsibility for the originality and correctness of all AI-assisted content.

\paragraph{Acknowledgements.} I thank Christian Coester for introducing the problem to me and for many insightful discussions on the problem.

\printbibliography
\appendix
\section[Proof of the Electrical Characterization]{Proof of \cref{lem:flow}}\label{app:electrical}
\begin{proof}
A feasible flow exists by sending positive excesses to negative ones. As $F$ varies over feasible flows, the vectors $(\sqrt{d_{ij}}F_{ij})_{i<j}$ form an affine subspace of a finite-dimensional Euclidean space. Such a subspace is closed and has a point of minimum norm. This proves attainment, even when some $d_{ij}=0$.

At a minimum, sending a small additional amount around a directed triangle $i,j,r$ preserves the net amount at each label. The derivative of the cost in this direction must vanish, giving
\[
  d_{ij}F_{ij}+d_{jr}F_{jr}+d_{ri}F_{ri}=0.
\]
Fix a reference label $r$, put $h_r=0$, and set $h_i=d_{ir}F_{ir}$. The triangle identity gives $d_{ij}F_{ij}=h_i-h_j$ for every pair; pairs containing $r$ satisfy this by definition.

Conversely, suppose a feasible $F$ with excess vector $\vec{a}$ has these voltages. For any flow $G$ with excess vector $\vec{b}$, grouping terms by labels gives
\[
  \sum_{i<j}d_{ij}F_{ij}G_{ij} =\sum_{i<j}(h_i-h_j)G_{ij} =\sum_i h_i\sum_jG_{ij}=\sum_i h_i b_i.
\]
Expanding the square of $F+G$ therefore yields
\[
  \sum_{i<j}d_{ij}(F_{ij}+G_{ij})^2 =\sum_{i<j}d_{ij}F_{ij}^2+2\sum_i h_i b_i+\sum_{i<j}d_{ij}G_{ij}^2.
\]
If $\vec{b}=\vec{0}$, this shows that changing a feasible flow without changing its net amounts cannot lower the energy, so $F$ is minimizing. For general $\vec{b}$, adding $F$ is a bijection from flows with excess $\vec{b}$ to flows with excess $\vec{a}+\vec{b}$. Taking minima proves the quadratic expansion in \cref{lem:flow}.
\end{proof}

\section{The Star Example}\label{app:star}
Take $k=8$, with the adversary locations at the center and seven leaves of the star in \cref{fig:star-energy}. The center-to-leaf distances are $1$ and the distances between distinct leaves are $2$. The center $s$ is empty, one leaf $a$ has two online servers, and each other leaf has one.

The initial energy is $1/3$. Serving from $a$ balances the counts and leaves energy $0$; serving from any other leaf leaves energy $4/9$. All eight servers are at distance $1$ from the center, so the expected decrease is
\[
  \E[\text{energy decrease}] =\frac28\left(\frac13-0\right)+\frac68\left(\frac13-\frac49\right)=0,
\]
while every move costs $1$. Hence no positive scaling of energy alone pays for this request.

\paragraph{Voltage Calculations.}
For the initial excess $\vec{e}_a-\vec{e}_s$ in \cref{fig:star-energy}, take voltages $1/3$ at leaf $a$, $1/9$ at every other leaf, and $0$ at the center $s$. For excess $\vec{e}_a-\vec{e}_i$, with $i$ another leaf, take voltages $2/9$ at $a$, $-2/9$ at $i$, and $0$ elsewhere. The voltage equations and net flows verify that these are minimizing flows, with energies $1/3$ and $4/9$ respectively.

For the center's spreading excess $\vec{e}_s-\one/8$, use voltage $1/8$ at the center and $0$ at the leaves. For a leaf's spreading excess, use voltage $2/9$ at that leaf, $1/72$ at the center, and $0$ elsewhere. Their energies are
\[
  c_s=\frac7{64},\qquad c_i=\frac{37}{192}\quad(i\ne s),\qquad c_i-c_s=\frac1{12}.
\]

\end{document}